\documentstyle[12pt]{article}
\begin{document}
\begin{center}
{\large \bf Spinor formulation of topologically massive gravity}\\[15mm]
{\large A. N. Aliev  and  Y. Nutku}\\[2mm]
 T\"{U}B\.{I}TAK - Marmara Research Center \\
 Research Institute for Basic Sciences \\
 Department of Physics \\
 41470 Gebze, Turkey \\[10mm]
\end{center}
\noindent

\vspace{1cm}

   In the framework of real 2-component spinors in three dimensional
space-time we present a description of topologically massive gravity
(TMG) in terms of differential forms with triad scalar coefficients.
This is essentially a real version of the Newman-Penrose
formalism in general relativity.
A triad formulation of TMG was considered earlier by Hall,
Morgan and Perjes, however, due to an unfortunate choice
of signature some of the spinors underlying the Hall-Morgan-Perjes
formalism are real, while others are pure imaginary.
We obtain the basic geometrical identities as well as the TMG field
equations including a cosmological constant for the appropriate signature.
As an application of this formalism we discuss the Bianchi Type
$VIII - IX$ exact solutions of TMG and point out that they are
parallelizable manifolds. We also consider various re-identifications
of these homogeneous spaces that result in black hole solutions of TMG.

\vspace{1cm}

\section{Introduction}

   The interesting properties of Einstein's theory of gravitation
in three dimensions are well-recognized
\cite{witten,gak,djth}. Due to the equivalence between
the Riemann and Ricci tensors in three dimensions, the space-time
curvature is completely determined by the energy-momentum tensor of the
exterior matter sources. In the absence of sources
space-time is flat, while the coupling to particle-like sources
\cite{djth} as well as the presence of a cosmological
constant \cite{dj,btz} provide exact
solutions with non-trivial static geometry. 
On the other hand the topology can be non-trivial even for vacuum
\cite{witten}.

   There is, however, a dynamical theory of gravity in three dimensions.
This is Deser, Jackiw and Templeton's theory \cite{djt} of topologically
massive gravity (TMG). The geometry of exact
solutions of TMG is non-trivial and in order to investigate the properties
of these exact solutions it will be useful to develope a triad formalism
analogous to the Newman-Penrose formalism \cite{np,ghp} of general relativity
and relate it to real 2-component spinors.
Earlier Hall, Morgan and Perjes \cite{hmp} (HMP) have presented
a triad formalism  for vacuum TMG without explicit reference to spinors.
Due to an unfortunate choice of signature some of the 2-component
spinors underlying HMP formalism are real while others are pure
imaginary. In view of the fundamental importance of a triad formalism
based on real spinors for TMG, we shall obtain the basic geometrical
identities as well as the field equations of TMG with a cosmological constant
for the appropriate signature and relate it to real 2-component spinors.
We shall express these results in terms of differential forms with
triad scalar coefficients. This is based on some unpublished work by one of
us \cite{old} for the Newman-Penrose formalism in general relativity.
The formulation of TMG in terms of differential forms was
considered earlier by Dereli and Tucker \cite{dt} without the use
of triad scalars.

    The Bianchi Type $VIII$ and $IX$ exact solutions of TMG \cite{nb}
are obtained readily through the use of this spinor formalism and therefore
they provide a simple illustration of our approach.
We shall show that the Bianchi Type $VIII$ solution can be re-identified,
{\it \'{a} l\`{a}} Ba\~{n}ados, Teitelboim and Zanelli \cite{btz},
to result in a generalization of the black hole solution of TMG \cite{n}.
Finally we shall show that Bianchi Type $VIII - IX$ exact solutions of TMG
are oriented, parallelizable manifolds which provide new examples
of 3-manifolds where the ideas of Cartan and Schouten are
realized \cite{cartsch}.

\section{Spin frame and the triad}

   We shall consider a 3-dimensional Riemannian space-time
described by a metric with
signature $  (\,+\,\,    - \,\, - \,) $.  In this case we can
introduce real 2-component spinors in complete analogy with
the spinor representation of 4-dimensional space-time \cite{np}.
For this purpose we start by introducing a spin frame with the basis
\begin{equation}
\zeta^{A}_{\,(a)} =
\{ \, o^{A} \, ,  \;\; \iota^{A} \, \} \;\;\;\;\;\;\; A = 1, 2
\;\;\; a = 0, 1
\label{01}
\end{equation}
of real 2-component spinors where, capital Latin indices refer to
the components of spinors and small Latin indices enclosed by
parantheses run over the two values omicron and iota respectively.
These indices will be raised and lowered by the Levi-Civita
symbol $\epsilon_{AB}$ from the right. Thus
\begin{equation}
o_A = o^B \epsilon_{BA} = - \epsilon_{AB} o^B
\label{minus}
\end{equation}
and the normalization of the frame will be determined by
\begin{equation}
o_A \iota^A  = 1
\label{01contr}
\end{equation}
with all others vanishing identically.
The triad of real vectors that corresponds to such a spin frame
will be given in terms of the basis spinors through
the Infeld-van der Waerden connecting quantities $ \sigma^{i}_{\;AB} $
which are symmetric $ \sigma^{i}_{\;AB} = \sigma^{i}_{\;BA} $.
We have
\begin{equation}     \begin{array}{cllll}
l^i & \equiv & \sigma^{i}_{\;AB}   o^A  o^B & = & \sigma^{i}_{\;00}
\\[2mm]
n^i & \equiv & \sigma^{i}_{\;AB}   \iota^A \iota^B & = & \sigma^{i}_{\;11}
\\[2mm]
 \frac{1}{\sqrt{2}}  m^i & \equiv & \sigma^{i}_{\;AB}  o^A \iota^B &
= & \sigma^{i}_{\;01}
= \sigma^{i}_{\;10}
\end{array}         \label{deftr}
\end{equation}
where $ l, n $ are null and $m$ is space-like.
Corresponding to the normalization in eq.(\ref{01contr})
they will be subject to
\begin{equation}
           l_i n^i =  l^i n_i =  1  \, , \;\;\;\;\; m_i m^i = - 1
\label{norm}
\end{equation}
with all other contractions vanishing identically. In compact notation
\begin{equation}
\sigma^{i}_{\;AB}      =   \left(       \begin{array}{cc}
l^i  &  \frac{1}{\sqrt{2}} m^i  \\[1mm]   \frac{1}{\sqrt{2}}  m^i & n^i        \end{array}             \right)
\label{sigup}
\end{equation}
and its inverse is given by
\begin{equation}
\sigma_{i}^{\;AB}      =   \left(       \begin{array}{cc}
n_i  &  - \frac{1}{\sqrt{2}}  m_i  \\ - \frac{1}{\sqrt{2}} m_i & l_i
\end{array}             \right)  .
\label{sigdown}
\end{equation}
As a consequence of the definitions (\ref{sigup}) and (\ref{sigdown})
we have the completeness relations
\begin{equation}
\delta^{i}_{\;k} =  \sigma^i_{\;AB} \sigma_k^{\;AB} =
l^i n_k + n^i l_k - m^i m_k \, ,
\label{complete}
\end{equation}
\begin{equation}
 \sigma^i_{\;MN} \sigma_i^{\;AB} = \frac{1}{2} \left(
 \delta^{A}_{M}  \delta^{B}_{N} +  \delta^{A}_{N}  \delta^{B}_{M} \right)
\label{complete2}
\end{equation}
and we recall the basic spinor identity in two dimensions
\begin{equation}
 \epsilon_{M \, [ \, A} \; \epsilon_{B\,C \, ] } = 0
\label{basicid}
\end{equation}
where square parantheses denote skew symmetrization. One consequence of
the identity (\ref{basicid}) which will be used repeatedly is
\begin{equation}
 \xi_A \eta_B  - \xi_B \eta_A  = \epsilon_{AB} \xi_M \eta^M
\label{basicid2}
\end{equation}
for any two 2-component spinors $ \xi$ and $ \eta$.

    The components of the metric are given by
\begin{equation}
g_{ik} =  \sigma_i^{\;AB}  \sigma_k^{\;MN} \epsilon_{AM} \epsilon_{BN}
 = l_i n_k + n_i l_k - m_i m_k
\label{metcomp}
\end{equation}
and using eq.(\ref{sigdown}) we may introduce the co-frame
\begin{equation}
\sigma_{i}^{\;AB} \, d x^i     =   \left(       \begin{array}{cc}
n  &  - \frac{1}{\sqrt{2}} m  \\ - \frac{1}{\sqrt{2}} m & l
\end{array}             \right)
\label{coframe}
\end{equation}
so that the basis 1-forms are given in terms of the triad
\begin{equation}
\{ n, l, - m \}  = e^{a}_{\;i} d x^{i} \;\;\;\;\;
\end{equation}
where,
\begin{equation}
e^{a}_{\;i} = \{ n_i,\, l_i,\, - m_i  \}
\label{triadup}
\end{equation}
is the inverse of the triad basis
\begin{equation}
e^{i}_{\;a} = \{ l^i,\, n^i,  m^i \} .
\label{triaddown}
\end{equation}
The triad components of the metric are
\begin{equation}
g_{ab} =  \left(       \begin{array}{ccc}
0 & 1 & 0 \\ 1 & 0 & 0 \\ 0 & 0 & - 1
\end{array}             \right)
\label{triadcom}
\end{equation}
while the space-time interval expressed in terms of
the basis 1-forms is given by
\begin{equation}
 d s^{\;2} = l \otimes n + n \otimes l - m \otimes m .
\label{metric}
\end{equation}
% The raising and lowering of the triad indices will be done
% with the metric  (\ref{triadcom}).
Next we define the intrinsic derivative operators, or the derivative
in the direction of the legs of the triad through
\begin{equation}
D = l^i  \frac{\partial}{\partial x^i} \;\;\;\;\;
\Delta = n^i  \frac{\partial}{\partial x^i} \;\;\;\;\;
\delta = m^i  \frac{\partial}{\partial x^i}
\label{covtriad}
\end{equation}
and
\begin{equation}
 d = l \, \Delta + n \, D -  m \, \delta
\label{d}
\end{equation}
is the resolution of the exterior derivative along the triad.

\section{Ricci rotation coefficients}

    Starting with the triad basis we have the Ricci rotation coefficients
which are the components of the Levi-Civita connection
\begin{equation}                  \begin{array}{lll}
   \gamma_{cab} & = & e_{a \; i; k} \, e^{i}_{\;c}  e^{k}_{\;b} \\
     \gamma_{cab} & = &  - \gamma_{acb}
\end{array}
\label{defricrot}
\end{equation}
where semicolon denotes the covariant derivative.
Using definitions analogous to those in the Newman-Penrose formalism,
the Ricci rotation coefficients can be named
\begin{equation}                  \begin{array}{lllll}
\gamma_{100}   & = &  \epsilon    &  = & l_{i ; k } \, n^{i }  l^{k } \\
\gamma_{101}   & = &  - \epsilon' &  = & l_{i ; k } \, n^{i }  n^{k } \\
\gamma_{102}   & = &  \alpha      &  = & l_{i ; k } \, n^{i }  m^{k } \\
\gamma_{200}   & = &  \kappa      &  = & l_{i ; k } \, m^{i }  l^{k } \\
\gamma_{201}   & = &  \tau        &  = & l_{i ; k } \, m^{i }  n^{k } \\
\gamma_{202}   & = &  \sigma      &  = & l_{i ; k } \, m^{i }  m^{k } \\
\gamma_{210}   & = &  \tau'       &  = & n_{i ; k } \, m^{i }  l^{k } \\
\gamma_{211}   & = &  \kappa'     &  = & n_{i ; k } \, m^{i }  n^{k } \\
\gamma_{212}   & = &  \sigma'     &  = & n_{i ; k } \, m^{i }  m^{k }
\end{array}
\label{rotcoeff}
\end{equation}
where prime denotes the analogue of the prime symmetry operation of
Geroch, Held and Penrose \cite{ghp} which results from
the interchange of $l$ and $n$ leaving $m$ fixed.
This operation does not change the
normalization conditions in eqs.(\ref{01contr}), (\ref{norm})
and we note that $ \alpha' = - \alpha $.
The Ricci rotation coefficients can be grouped together in a table
\begin{center}
\begin{tabular}{|cc|c|c|c|}                \hline
    & $ca$   &   10  &  20   & 21  \\
$b$      &        &      &  &    \\ \hline

0   & &  $\epsilon$ &   $\kappa$ &   $\tau'$ \\
    & &            &            &       \\
1   & &  $ - \epsilon'$ &$ \tau$ & $\kappa'$  \\
    & &            &             &          \\
2   & &    $\alpha$ & $\sigma$  & $\sigma'$     \\
    &  &          &                &              \\ \hline
\end{tabular}
\end{center}

\begin{center}
Table 1.
\end{center}

    One of the advantages of presenting the triad formalism of
TMG in terms of differential forms lies in the great simplification
it provides for the calculation of the Ricci rotation coefficients.
Taking the exterior derivative of the basis 1-forms and
expressing the result in terms of the basis 2-forms
yields the spin coefficients through a linear algebraic system.
That is, by a comparison of the results with
\begin{equation}                  \begin{array}{cll}
d \, l & = & - \epsilon \; l \wedge n +  (\alpha - \tau ) \;
           l \wedge m  - \kappa \; n \wedge m  \\[2mm]
d n & = &  \epsilon' \; l \wedge n  - \kappa' \; l \wedge m
- ( \alpha  + \tau' ) \; n \wedge m  \\[2mm]
d  m & = & ( \tau' - \tau ) \; l \wedge n - \sigma' \;
l  \wedge m  - \sigma \; n  \wedge m
\end{array}
\label{dlnm}
\end{equation}
we can obtain the Ricci rotation coefficients through the solution of
a linear algebraic system.

\section{Spin coefficients}

    Turning to the definition of spin coefficients
which are the components of the spin connection we
shall begin with the spin frame and use
\begin{equation}
\Gamma_{(a)(b)(c)(d)} = \zeta_{(a)\,A;\,CD}  \zeta_{(b)}^A
  \zeta_{(c)}^C  \zeta_{(d)}^D ,
\end{equation}
or more briefly
\begin{equation}
\Gamma_{(a)(b)\,CD} = \zeta_{(a)\,A;\,CD}  \zeta_{(b)}^A
\end{equation}
where, once again, semicolon stands for the covariant derivative.
The correspondence between the covariant derivatives of the vector and
spinor fields is established by means of the Infeld-van der Waerden
connecting quantities (\ref{sigup}) and (\ref{sigdown}). Thus
\begin{equation}    \begin{array}{rcl}
\Psi_{i ;\, k} & = & \sigma_{i }^{AB} \, \sigma_{k }^{CD} \Psi_{AB;CD}\\[2mm]
\Psi_{AB;CD} & = & \sigma_{AB}^{i} \sigma_{CD}^{k} \Psi_{i;k}   
\end{array}
\end{equation}
as a consequence of which we can easily verify that
\begin{equation}
\sigma_{AB;\,CD}^{i}=0, \hspace{1cm} \sigma_{i\hspace{5mm};CD}^{AB}=0
\hspace{1cm} \epsilon_{AB\,;\;CD} = 0 .
\end{equation}
The spin coefficients
are symmetric  both in the first and the last pairs of indices
\begin{equation}
\Gamma_{(a)(b)(c)(d)}=\Gamma_{(b)(a)(c)(d)}=\Gamma_{(a)(b)(d)(c)}
\label{symind}
\end{equation}
since the spin frame is defined solely through real spinors.
The definition of spin coefficients can be given through the formula
\begin{equation}
\Gamma_{(a)(b)\,XY}= \frac{1}{2}\, \epsilon^{(c)(d)} \zeta_{(a)}^A
 \zeta_{(d)}^D \left [\zeta_{(b)A} \zeta_{(c)D} \right]_{;XY}
\label{praccalfor}
\end{equation}
which is a real version of the analogous formula
in general relativity  \cite{w}.
This can be verified by direct calculation
\begin{eqnarray}
\Gamma_{(a)(b)\,XY} & = & \frac{1}{2}\, \epsilon^{(c)(d)}\left[ \zeta_{(a)}^A
\zeta_{(b)A} \zeta_{(d)}^D \zeta_{(c)D;XY}+ \zeta_{(a)}^A \zeta_{(d)}^D
\zeta_{(c)D} \zeta_{(b)A\,;XY}\right] \nonumber \\
& = &\frac{1}{2}\,\epsilon^{(c)(d)}
\left[-\epsilon_{(a)(b)}\,\zeta_{(d)}^D \zeta_{(k)D} {\Gamma^{(k)}}_{(c)XY}
+\epsilon_{(c)(d)} \zeta_{(a)}^A \zeta_{(k)A} {\Gamma^{(k)}}_{(b)XY}\right]
\nonumber \\
& = &
\frac{1}{2}\,\epsilon^{(c)(d)}\epsilon_{(c)(d)}\delta^{(k)}_{(a)}
\Gamma_{(k)(b)XY}=\Gamma_{(a)(b)XY} .
\label{spcoeff1}
\end{eqnarray}
Using eq.(\ref{praccalfor}) we can evaluate the spin coefficients
in terms of the Ricci rotation coefficients. As an example we have
\begin{eqnarray}
\Gamma_{0000} & = & \frac{1}{2}\left[  o^A \iota^B \left( o_{A} o_{B}
\right)_{;00} - o^A o^B \left( o_{A} \iota_{B}\right)_{;00}\right]
\nonumber \\
& = &
\frac{1}{2\sqrt{2}} \left( l_{i;k}\,l^{k} m^{i} - m_{i;k} l^{i}l^{k}
\right) = \frac{1}{ \sqrt{2}}\, \kappa  \label{spcoeff2}
\end{eqnarray}
% \begin{eqnarray}
% \Gamma_{1111} & = & \frac{1}{2}\left[ \iota^{A} \iota^{B}\left( \iota_{A}
% o_{B}\right)_{;11} - \iota^{A}o^B \left( \iota_{A} \iota_{B} \right)_{;11}
% \right] \nonumber \\
% & =&
% \frac{1}{2\sqrt{2}} \left( m_{i;k} n^{i}n^{k} -n_{i;k} n^{k}m^{i}\right)
% =- \frac{1}{\sqrt{2}}\,\kappa'
% \end{eqnarray}
and the rest of the spin coefficients can be evaluated similarly.
It is clear that due to their symmetry properties (\ref{symind})
only nine spin coefficients need to be evaluated.
% Taking into account
%the relation  $ \Gamma_{abcd} = \Gamma^{\,\,q}_{a \,\,\,cd}\,\epsilon_{qb} $
%we find that the  quantities   $  \Gamma^{\,\,b}_{a \,\,\,cd} $ can be listed
% according to the table
We find that the  spin coefficients can be listed
according to the table
\begin{center}
\begin{tabular}{|ccc|c|c|c|}           \hline
      &      & $   b $  &  0 &  1 &  0 \\
      & $a$  &          & 0  & 0  & 1 \\
$ cd $&      &          &    &    &   \\       \hline
      &      &          &    &    &    \\
00    & &          &  $ \frac{1}{2} \epsilon $ &
$ -  \frac{1}{\sqrt{2}} \kappa $  &
 $ -  \frac{1}{\sqrt{2}} \tau' $    \\
      & &           &              &           &  \\
01  &   &  &$  \frac{1}{2 \sqrt{2}} \alpha $   &
 $  - \frac{1}{2} \sigma  $   &  $   - \frac{1}{2} \sigma' $   \\
    &   &  &                  &                &                \\
11  & &  &  $  -  \frac{1}{2} \epsilon' $   &
$ - \frac{1}{\sqrt{2}} \tau  $  &
  $  - \frac{1}{\sqrt{2}} \kappa' $    \\
    & &    &                   &            &    \\    \hline
\end{tabular}
\end{center}
\begin{center}
 Table 2.
\end{center}
where $ \Gamma_{abcd} = \Gamma^{\,\,q}_{a \,\,\,cd}\,\epsilon_{qb} $.

\section{Connection 1-form}

  The connection 1-form is defined by
\begin{equation}  \Gamma_{a}^{\;\;b}      =
\Gamma_{a\;\;cd}^{\;\;b}\; \sigma_{i}^{cd} d x^i
\label{pot}
\end{equation}
which is a traceless  $ 2 \times 2 $ real matrix of 1-forms and
using eq.(\ref{sigdown}) and table $ 2 $ we have
\begin{equation} \begin{array}{llclc}
\Gamma_{0}^{\;\;0} & = &  \frac{1}{2}  l_{i;k} n^i d x^k    &  = &
\frac{1}{2}    \left(
- \epsilon' \, l +  \epsilon \, n - \alpha \, m    \right) \\[2mm]
\Gamma_{0}^{\;\;1}  &  =  &  - \frac{1}{\sqrt{2}} l_{i;k} m^i d x^k
& = &  \frac{1}{\sqrt{2}}   \left(
  -  \tau \, l - \kappa \, n + \sigma \, m    \right)  \\[2mm]
\Gamma_{1}^{\;\;0}  &  =  &  - \frac{1}{\sqrt{2}}  n_{i;k} m^i d x^k  & = &
  \frac{1}{\sqrt{2}}   \left(
- \kappa' \, l -  \tau' \, n + \sigma'  \, m \right)
\end{array}                \label{potform}
\end{equation}
with  $  \Gamma_{1}^{\;\;1} = - \Gamma_{0}^{\;\;0}  $.
Under the prime operation the entries of  $  \Gamma_{a}^{\;\;b} $
flip, $  \left(\Gamma_{0}^{\;\;0} \right)' =
 \Gamma_{1}^{\;\;1} , \;\;\;\;\;  \left(\Gamma_{1}^{\;\;0} \right)' =
 \Gamma_{0}^{\;\;1}  $ and for the rest we recall that
the prime operation is an involution. The most prominent role
played by the connection 1-form is in holonomy
\begin{equation}
P \, e^{ i \int_{- \infty}^x  \Gamma_{a}^{\;\;b} }
\end{equation}
where $P$ denotes path-ordering.

\section{Curvature 2-form}

   The curvature 2-form is obtained from the connection 1-form through
\begin{equation}
  R_{a}^{\;\;b} = d \, \Gamma_{a}^{\;\;b} -  \Gamma_{a}^{\;\;m}
\wedge    \Gamma_{m}^{\;\;b}
\label{curv}
\end{equation}
but first we must define the triad scalars for curvature.
In three dimensions there is no Weyl tensor
and the Riemannian curvature tensor can be decomposed in terms
of the Ricci tensor $  R_{ab} =  R^{c}_{\;\; acb}  $ and
the scalar of curvature $R$. Using the traceless Ricci tensor
\begin{equation}
\Theta_{ab} = - \frac{1}{2} \left( R_{ab} -
 \frac{1}{3} g_{ab} R \right)
\label{trricci}
\end{equation}
the Riemann tensor is given by \cite{e}
\begin{equation}
R_{abcd} = - 2 g_{ac} \Theta_{bd} + 2 g_{ad} \Theta_{bc}
- 2   g_{bd} \Theta_{ac}
+ 2 g_{bc} \Theta_{ad} -  \frac{1}{6}  ( g_{ad} g_{bc} - g_{ac} g_{bd} ) R .
\label{riemricci}
\end{equation}
and we can pass to the spinor analogue of this expression
\begin{equation} \begin{array}{lll}
R_{AWBXCYDZ} & = & - 2 \Phi_{BDXZ} \; \epsilon_{AC} \; \epsilon_{WY}
- 2 \Phi_{ACWY} \; \epsilon_{BD} \; \epsilon_{XZ}    \label{spincurv} \\[1mm]
  & &   + 2 \Phi_{BCXY} \; \epsilon_{AD} \; \epsilon_{WZ}
     + 2 \Phi_{ADWZ} \; \epsilon_{BC} \; \epsilon_{XY} \\[1mm]
   & &   + \frac{1}{6} \left(
 \epsilon_{AC} \; \epsilon_{WY}  \; \epsilon_{BD} \; \epsilon_{XZ}
 -  \epsilon_{AD} \; \epsilon_{BC}  \; \epsilon_{WZ} \;
         \epsilon_{XY} \right) R
\end{array}
\end{equation}
where the quantities   $ \Phi_{AWBX}  $ are  curvature spinors.
The spinor analogue of the trace-free Ricci tensor is similarly
\begin{equation}
R_{AWBX} = - 2 \Phi_{ABWX}  + 6 \Lambda \, \epsilon_{AB} \, \epsilon_{WX}
\label{trspin}
\end{equation}
where we have introduced the definition
\begin{equation}
 \Lambda = \frac{1}{18} R
\end{equation}
for the curvature scalar. With these preliminaries
the curvature 2-form can be written in the form
\begin{equation}
R_{m}^{\;\;n}=\frac{1}{2}\; \zeta^{M}_{(m)}  \zeta_{N}^{(n)}
 R_{MX\;\;\;\;CYDZ}^{\;\;\;\;\;\;NX}   \,
   \sigma_{i}^{CY}\,\sigma_{k}^{DZ}\,d x^i\wedge dx^k
\label{curv2}
\end{equation}
and in terms of the basis 2-forms we get
$$ R_{0}^{\;\;0}      =
 \left( 2 \Phi_{11} - \frac{3}{2} \Lambda \right)\, l \wedge n
- \Phi_{12}\, l \wedge m + \Phi_{10}\, n \wedge m $$
\begin{equation}
R_{0}^{\;\;1}      =
-  \sqrt{2} \, \Phi_{01}\, l \wedge n  + \frac{1}{\sqrt{2}}\, \Phi_{02}\,
l \wedge m - \sqrt{2}\, \Phi_{00}\, n \wedge m
\label{curv23}
\end{equation}
\vspace{0.2 mm}
$$ R_{1}^{\;\;0}      =
  \sqrt{2}\,  \Phi_{12}\, l \wedge n - \sqrt{2}\, \Phi_{22}\, l \wedge m
 + \frac{1}{\sqrt{2}} \, \Phi_{02}\, n \wedge m   $$
where we have introduced the definitions
\begin{equation}                       \begin{array}{llllllllll}
 \Phi_{00} & := & \Phi_{0000}, \hspace{5mm}      &
 \Phi_{01} & := & \sqrt{2}\, \Phi_{0010},  \hspace{5mm}        &
 \Phi_{02} & := & 2\, \Phi_{0011},  \hspace{5mm}           \\ [2mm]
 \Phi_{11} & := & \Phi_{0101},      &
 \Phi_{12} & := & \sqrt{2}\,\Phi_{0111},      &
 \Phi_{22} & := & \Phi_{1111}           .
 \end{array}
\end{equation}
Taking into account the equation (\ref{trspin}) along with the relations
(\ref{sigup}) and  (\ref{sigdown}) we see that these quantities
are triad scalars
\begin{equation}                  \begin{array}{lllllllll}
\Phi_{00}  & = & - \frac{1}{2} R_{ik} \, l^i l^k ,     &
\Phi_{01}  & = & - \frac{1}{2} R_{ik} \, l^i m^k  ,    &
\Phi_{02}  & = & - \frac{1}{2} R_{ik} \, m^i m^k   ,    \\[2mm]
\Phi_{11}  & = & - \frac{1}{2} R_{ik} \, l^i n^k + 3 \Lambda , &
\Phi_{12}  & = & - \frac{1}{2} R_{ik} \, n^i m^k ,  &
\Phi_{22}  & = & - \frac{1}{2} R_{ik} \, n^i n^k .
\end{array}
\label{curvcoeff}
\end{equation}
Note that   $  \Phi_{ik} $ is symmetric and  under the prime operation
the index 1 remains unchanged, while the rest flip
$ 0 \leftrightarrow 2 $.

\section{Ricci and Bianchi identities}

  From the definition of the curvature 2-form (\ref{curv}) and
eqs.(\ref{potform}), (\ref{curv23}) as well as the relations in
eqs.(\ref{d}), (\ref{dlnm}) we obtain 
\begin{equation}
D \epsilon' + \Delta \epsilon + 2  \epsilon \epsilon'
+ \alpha ( \tau -  \tau' ) - \tau \tau' + \kappa \kappa'
 = 4 \Phi_{11} - 3 \Lambda
\label{ricci1}
\end{equation}
\begin{equation} \begin{array}{rcl}
\delta \kappa - D \sigma + \sigma^2 - \kappa ( \tau + \tau' )
+ \sigma  \epsilon - 2  \kappa \alpha & = & 2 \Phi_{00} \\
& & \\
\delta \kappa' - \Delta \sigma' + \sigma'^2 - \kappa' ( \tau + \tau' )
+ \sigma'  \epsilon' + 2  \kappa' \alpha & = & 2 \Phi_{22}
\end{array}
\label{ricci2}
\end{equation}
\begin{eqnarray}
 \Delta \sigma - \delta \tau
- \sigma \sigma' +  \kappa \kappa' + \epsilon' \sigma
+ \tau^2 & = & \Phi_{02}  \nonumber \\ \label{ricci3}
& & \\
D \sigma' - \delta \tau'
- \sigma \sigma' +  \kappa \kappa' + \epsilon  \sigma'
 + \tau'^2 & = & \Phi_{02}  \nonumber
\end{eqnarray}
\begin{eqnarray}
 \Delta \alpha + \delta \epsilon' + \epsilon' ( \alpha - \tau )
- \sigma' ( \alpha + \tau ) + \kappa' ( \epsilon + \sigma ) & =& 2 \Phi_{12}
\nonumber  \\                          \label{ricci4}
& & \\
- D \alpha + \delta \epsilon - \epsilon ( \alpha + \tau' )
+ \sigma ( \alpha - \tau' ) + \kappa ( \epsilon' + \sigma' ) & = & 2 \Phi_{01}
\nonumber
\end{eqnarray}
\begin{eqnarray}
\Delta \kappa - D \tau + 2 \kappa \epsilon' + \sigma ( \tau - \tau' )
& = & 2 \Phi_{01} \nonumber \\         \label{ricci5}
& &  \\
D \kappa' - \Delta \tau' + 2 \kappa' \epsilon - \sigma' ( \tau - \tau' )
& = & 2 \Phi_{12}  \nonumber
\end{eqnarray}
which make up the set of Ricci identities.

  The full Bianchi identities are given by
\begin{equation}
d R_{a}^{\;\;b}  +  \Gamma_{a}^{\;\;m} \wedge R_{m}^{\;\;b}
               -  R_{a}^{\;\;m} \wedge \Gamma_{m}^{\;\;b} = 0
\label{bianchi}
\end{equation}
and we recall that in 3 dimensions
there is no difference between the full and contracted Bianchi identities.
Using the same procedure that we have used to obtain the Ricci identities
we can write the Bianchi identities (\ref{bianchi}) in terms of triad
scalars
\begin{equation}  \begin{array}{rcl}
 \Delta \Phi_{01} + D \Phi_{12}
- 2 \delta ( \Phi_{11} - \frac{3}{4} \Lambda )
+ 2 ( \tau + \tau' ) ( \Phi_{11} - \frac{3}{4} \Lambda )
+ \kappa \Phi_{22} & & \\  + \kappa' \Phi_{00}
- ( 2 \sigma - \epsilon )  \Phi_{12}   
- ( 2 \sigma' - \epsilon' ) \Phi_{01}   
+ \frac{1}{2} ( \tau + \tau' ) \Phi_{02}
 &  = & 0
\end{array}
\label{bianchi1}
\end{equation}
\begin{equation}  \begin{array}{rcl}
 D \Phi_{02} + 2 \Delta \Phi_{00}  - 2 \delta \Phi_{01}
+ 2 ( 2 \tau + \tau' + \alpha ) \Phi_{01}
-  \sigma \Phi_{02}
& &      \\ - 2 ( \sigma' - 2 \epsilon' ) \Phi_{00}
+ 2 \kappa \Phi_{12}
- \sigma \left( 4 \Phi_{11} - 3 \Lambda \right)   & =  &  0  \\[2mm]
 \Delta \Phi_{02} + 2 D \Phi_{22} - 2 \delta \Phi_{12}
+ 2 ( 2 \tau' + \tau - \alpha )  \Phi_{12}
-  \sigma' \Phi_{02}   & &    \\
- 2 (  \sigma - 2 \epsilon ) \Phi_{22}
+ 2 \kappa' \Phi_{01}
- \sigma' \left( 4 \Phi_{11} - 3 \Lambda \right) & = & 0
\end{array}
\label{bianchi2}
\end{equation}
Under the prime operation the unpaired Ricci and Bianchi identities
(\ref{ricci1}), (\ref{bianchi1}) remain invariant
while the paired equations (\ref{ricci2}),
(\ref{ricci3}), (\ref{ricci4}), (\ref{ricci5}), (\ref{bianchi2})
simply exchange places.

\section{Cotton 2-form}

   The conformal properties of 3-dimensional space-time are described
by the Cotton tensor \cite{e}
\begin{equation}
C^{ik} = \epsilon^{ijl} \left( R_{l}^{k} - \frac{1}{4} \, \delta_{l}^{k} R
\right)_{;j}
\end{equation}
where $ \epsilon^{ijl} $ is the completely skew 3-dimensional Levi-Civita
tensor density of weight $-1$. It remains invariant under conformal
transformations of the 3-dimensional metric and vanishes
for conformally flat metrics. The Cotton tensor satisfies the conditions
\begin{equation}
C^{ik} = C^{ki} , \; \;\;\;\;\;
  C^{i}_{\,i} = 0 , \; \;\;\;\;\;   C^{ik}_{\;\;\,;k} = 0
\end{equation}
of a symmetric, traceless, covariantly conserved tensor.

    For the purpose of writing the TMG field equations in terms of
differential forms we need to construct a 2-form out of the Cotton tensor.
The required expression is given by
\begin{equation}
C_{a}^{\;\;b} = d \, ^*{ \hspace{-0.7mm} } R_{a}^{\;\;b}
  - \Gamma_{a}^{\;\;m} \wedge \,  ^*{ \hspace{-0.7mm} }  R_{m}^{\;\;b} -  \,
^*{ \hspace{-0.7mm} } R_{a}^{\;\;m} \wedge \Gamma_{m}^{\;\;b} - \frac{1}{4}\,
dR \wedge\,\Sigma_{a}^{\;\;b}
\label{cotton}
\end{equation}
where the 1-form $  \Sigma_{a}^{\;\;b}  $  is obtained from the co-frame
(\ref{coframe}) by lowering an appropriate spinor index
\begin{equation}
\Sigma_{a}^{\;\;b}=\frac{1}{\sqrt{2}}\,\sigma_{a\;\;i}^{\;\;b}\,dx^i
\label{coframeform}
\end{equation}
and therefore we need the dual of the curvature 2-form
(\ref{curv2}) which in turn requires the duals of the basis 2-forms.
Starting with the definitions
\begin{equation}
^*{ \hspace{-0.5mm} } dx^{i}  =  \frac{1}{2} \, \epsilon^{ijk} dx^{j} \wedge
dx^{k}, \hspace{1cm} ^{* \,*} dx^{i}   =   dx^{i}
\end{equation}
and using the completeness relation (\ref{complete}) we find
\begin{equation}                  \begin{array}{ccc}
^*{ \hspace{-0.5mm} } \left(  l  \wedge n  \right) & = & - m \\
^*{ \hspace{-0.5mm} } \left(  l  \wedge m  \right) & = & - l \\
^*{ \hspace{-0.5mm} } \left(  n \wedge  m  \right) & = & n
\end{array}
\label{hodge}
\end{equation}
summarizing the effect of
the Hodge star operator on the basis 2-forms.
Then the duals of the curvature 2-forms (\ref{curv23}) are simply
$$ ^*{ \hspace{-0.7mm} }  R_{0}^{\;\;0} =  \Phi_{12}\, l  +  \Phi_{10}\, n
  - 2 \left( \Phi_{11} - \frac{3}{4} \Lambda \right)\, m  $$
\begin{equation}
^*{ \hspace{-0.7mm} }  R_{0}^{\;\;1} =
- \frac{1}{\sqrt{2}} \Phi_{02}\, l - \sqrt{2} \Phi_{00}\, n +
\sqrt{2} \Phi_{01}\, m
\label{dual}
\end{equation}
\vspace{4mm}
$$ ^*{ \hspace{-0.7mm} }  R_{1}^{\;\;0} =
  \sqrt{2} \Phi_{22}\, l + \frac{1}{\sqrt{2}} \Phi_{02}\, n  -
  \sqrt{2} \Phi_{12}\, m $$
and inserting into the equation (\ref{cotton}) the quantities (\ref{d}),
\, (\ref{potform}),\, (\ref{dual}) and (\ref{coframeform}) we can express
the Cotton 2-form in terms of the basis 2-forms. We find
\begin{equation} \begin{array}{lll}
C_{0}^{\;\;0} & = & \left\{   \Delta \Phi_{10}
 - D \Phi_{12}  + 3 ( \tau - \tau' ) \Phi_{11}
+ \epsilon' \Phi_{01} -  \epsilon \Phi_{12}
- \kappa \Phi_{22}  \right. \\[2mm]
& & \hspace{-1mm}   + \kappa' \Phi_{00}  \left.  \right\}  l \wedge n \;
+ \; \left\{ \delta \Phi_{12} - \Delta \Phi_{02} - 2 \kappa' \Phi_{01}
+ ( \alpha - 2 \tau ) \Phi_{12}
    \right. \\[2mm]  & &     \hspace{-1mm}
     + \sigma \Phi_{22}
   + 3\sigma' \Phi_{11} + \frac{9}{4}\Delta \Lambda
       \left.  \right\} l \wedge m    \;  + \;
\left\{ \delta \Phi_{01} - D \Phi_{02}  - 2 \kappa  \Phi_{12}
 \right. \\[2mm]   & & \hspace{-1mm}
- (  \alpha + 2 \tau' ) \Phi_{01}  + \sigma' \Phi_{00} + 3\sigma \Phi_{11}
+ \frac{9}{4} D \Lambda  \left.  \right\} n \wedge m
\end{array}                     \label{cotton1}
\end{equation}
\vspace{1mm}
\begin{equation} \begin{array}{lll}
 C_{0}^{\;\;1} & = & \sqrt{2} \left\{   D \Phi_{11} -  \Delta \Phi_{00}
 + ( \tau' -  2 \tau ) \Phi_{01} - 2  \epsilon' \Phi_{00} +
\kappa \Phi_{12}  \right. \\[2mm]
& & \hspace{-1mm}  -\frac{3}{4} D\Lambda  \left.  \right\}  l \wedge n \;
+ \; \sqrt{2} \left\{ \Delta \Phi_{01} - \delta \Phi_{11}
 + \kappa' \Phi_{00}  + 3 \tau \Phi_{11}       \right. \\[2mm]
& & \hspace{-1mm}  - \sigma \Phi_{12}
 + ( \epsilon' - \sigma' )  \Phi_{01} + \frac{3}{4}\delta
\Lambda   \left.  \right\}  l \wedge m \; + \;
\sqrt{2} \left\{  D \Phi_{01}  -  \delta \Phi_{00}   \right. \\[2mm]
& & \hspace{-1mm}    +3\kappa  \Phi_{11}
- ( \epsilon + 2 \sigma ) \Phi_{01}
+ ( \tau' + 2 \alpha )  \Phi_{00}  \left.  \right\}  n \wedge m
\end{array}                  \label{cotton2}
\end{equation}
\vspace{1mm}
\begin{equation} \begin{array}{lll}
 C_{1}^{\;\;0} & = & \sqrt{2} \left\{   \Delta \Phi_{11} -  D \Phi_{22}
 + ( \tau -  2 \tau' ) \Phi_{12} - 2  \epsilon \Phi_{22} +
\kappa' \Phi_{01}   \right. \\[2mm]
& & \hspace{-1mm} -\frac{3}{4} \Delta\Lambda  \left.  \right\}  l \wedge n \;
+ \;  \sqrt{2} \left\{ \delta \Phi_{22} - \Delta \Phi_{12}
  -3 \kappa' \Phi_{11} + ( \epsilon' + 2 \sigma' ) \Phi_{12}  \right. \\[2mm]
& & \hspace{-1mm}  - ( \tau -  2 \alpha ) \Phi_{22}  \left.  \right\}  l \wedge m  \;+  \;
 \sqrt{2} \left\{ \delta \Phi_{11} -  D \Phi_{12} + \sigma' \Phi_{01}
 - \kappa  \Phi_{22}    \right. \\[2mm]
& & \hspace{-1mm} + ( \sigma -  \epsilon ) \Phi_{12}
  -3 \tau' \Phi_{11}
- \frac{3}{4} \delta \Lambda  \left.  \right\}  n \wedge m
\end{array}                  \label{cotton3}
\end{equation}

Unlike the curvature 2-form, the cotton 2-form picks up a minus
sign under the prime operation in addition to the flip noted earlier
$ \left(  C_{0}^{\;\;1} \right)' = -  C_{1}^{\;\;0} $,
$ \left(  C_{0}^{\;\;0} \right)' = -  C_{1}^{\;\;1} $
which is due to its parity violating nature.

\section{Field equations}

   Deser, Jackiw and Templeton's theory \cite{djt} of TMG requires the
proportionality of the Einstein and Cotton tensors in 3-dimensions.
The field equations of TMG are given by
\begin{equation}
 G^{ik} + \frac{1}{\mu}\, C^{ik} = \lambda \, g^{ik}
\label{djt}
\end{equation}
where  $  G^{ik} $ is the Einstein tensor,
$  \mu  $ is the DJT constant of proportionality which can be regarded as
topological mass and  $  \lambda $ is the cosmological constant.
Using the curvature and Cotton 2-forms described in the previous sections,
we can rewrite the field equations (\ref{djt}) in terms of 2-forms
\begin{equation}
 R_{a}^{\;\;b} + \frac{1}{\mu} \, C_{a}^{\;\;b} +
 \lambda\, \Sigma_{a}^{\;\;m} \wedge \Sigma_{m}^{\;\;b}=0 .
\label{djt2}
\end{equation}
The expression of the field equations in terms of triad scalars
follows from the substitution of the results in eqs.(\ref{curv23}),
(\ref{coframeform}) and (\ref{cotton1}) - (\ref{cotton3})
into eq.(\ref{djt2}).

   The full set of TMG field equations are
\begin{equation} \begin{array}{rcl}
 D \Phi_{12}  - \Delta \Phi_{10}  - 3 ( \tau - \tau' ) \Phi_{11}
- \epsilon' \Phi_{01} +  \epsilon \Phi_{12} & &  \\
+ \kappa \Phi_{22}
- \kappa' \Phi_{00}
 &  = &  \mu ( \lambda  +  \Phi_{02} )
\label{tmg1}
\end{array}
\end{equation}
\vspace{2mm}
\begin{equation}   \begin{array}{rcl}
\delta \Phi_{01} - D \Phi_{02}  - 2 \kappa  \Phi_{12}
- (  \alpha + 2 \tau' ) \Phi_{01}
 + \sigma' \Phi_{00}           & &   \\
+3 \sigma \Phi_{11}+\frac{9}{4}D\Lambda
&  =  &  - \mu \Phi_{01}  \\[2mm]
  \delta \Phi_{12} - \Delta \Phi_{02}  - 2 \kappa'  \Phi_{01}
 + (  \alpha - 2 \tau ) \Phi_{12}
+ \sigma \Phi_{22}              &  &   \\
 +3 \sigma' \Phi_{11}
 +\frac{9}{4}\Delta\Lambda
& =  & \mu \Phi_{12}
\end{array}
\end{equation}
\vspace{3mm}
\begin{equation}   \begin{array}{rcl}
 D \Phi_{11} -  \Delta \Phi_{00}
 + ( \tau' -  2 \tau ) \Phi_{01}
 - 2  \epsilon' \Phi_{00} +
  \kappa \Phi_{12}-\frac{3}{4} D\Lambda
&  = &  \mu \Phi_{01}   \\[2mm]
\Delta \Phi_{11} -  D \Phi_{22} + ( \tau -  2 \tau' ) \Phi_{12 }
- 2 \epsilon \Phi_{22} + \kappa' \Phi_{01} -\frac{3}{4}\Delta\Lambda
& = &  - \mu \Phi_{12}
\end{array}
\end{equation}
\vspace{3mm}
\begin{equation}   \begin{array}{rcl}
 \Delta \Phi_{01} - \delta \Phi_{11} + \kappa' \Phi_{00}  + 3 \tau \Phi_{11}
 - \sigma \Phi_{12}  & &  \\
 + ( \epsilon' - \sigma' )  \Phi_{01} + \frac{3}{4}
\delta\Lambda
& = &   -\frac{1}{2} \mu ( \lambda  +  \Phi_{02} ) \\[2mm]
 D \Phi_{12} - \delta \Phi_{11} + \kappa \Phi_{22}  + 3 \tau' \Phi_{11}
 - \sigma' \Phi_{01}               & &  \\
 + ( \epsilon - \sigma )  \Phi_{12}
 +\frac{3}{4}\delta\Lambda
&  = &  \frac{1}{2} \mu ( \lambda  +  \Phi_{02} )
\end{array}
\end{equation}
\vspace{3mm}
\begin{equation}   \begin{array}{rcl}
  D \Phi_{01}  -  \delta \Phi_{00}
  +3 \kappa  \Phi_{11}  - ( \epsilon + 2 \sigma ) \Phi_{01}
+ ( \tau' + 2 \alpha )  \Phi_{00}
&  = &   \mu \Phi_{00}    \\[2mm]
\Delta \Phi_{12}  -  \delta \Phi_{22}
  +3 \kappa'  \Phi_{11}  - ( \epsilon' + 2 \sigma' ) \Phi_{12}
+ (  \tau - 2 \alpha )  \Phi_{22}
&  =  & - \mu \Phi_{22}
\end{array}
\end{equation}
Finally the DJT field equations (\ref{djt})  imply
\begin{equation}
\Lambda= - \frac{1}{3} \lambda
\label{lambda}
\end{equation}
since the Cotton tensor is traceless.

   We see that the TMG field equations remain under the prime
operation provided that the sign of the DJT coupling constant is also
changed simultaneously.

\section{Homogeneous Spaces}

   In order to illustrate the usefulness of the formalism we have
presented above we shall now consider its application to an
exact solution of the TMG field equations. This is a homogeneous space-time
related to the Bianchi Type $ VIII $ squashed 3-pseudo-sphere \cite{nb}.
We start with left-invariant 1-forms parametrized in terms of Euler angles
\begin{eqnarray}
\sigma^{0} & = & d\psi + \cosh{\theta} \, d\phi   \\
\sigma^{1} & = & - \sin{ \psi}\, d\theta
+  \cos{ \psi}\, \sinh{\theta} \, d\phi \\
\sigma^{2} & = & \cos{\psi}\,d\theta +\sin{\psi} \,\sinh{\theta} \, d\phi
\label{euler}
\end{eqnarray}
which are subject to the Maurer-Cartan equations
\begin{equation}
d\sigma^{\alpha}=\frac{1}{2}\, C_{\;\;\beta \gamma}^{\alpha}\,\sigma^{\beta}
\,\wedge\sigma^{\gamma}   \label{maurerc}
\end{equation}
with structure constants
\begin{equation}
 C_{\;\;12}^{0}= -1, \hspace{1cm}   C_{\;\;20}^{1}= 1, \hspace{1cm}
 C_{\;\;01}^{2}= 1
\label{struconst8}
\end{equation}
of Bianchi Type $VIII$.
In terms of these left-invariant 1-forms the TMG co-frame is defined by
\begin{equation}
\omega^{0}=\lambda_{0}\,\sigma^{0}, \hspace{1cm}
\omega^{1}=\lambda_{1}\,\sigma^{1}, \hspace{1cm}
\omega^{2}=\lambda_{2}\,\sigma^{2},
\label{homframe}
\end{equation}
where  $  \lambda_{0} , \, \, \,   \lambda_{1}, $   and  $  \lambda_{2}   $
are constants. The exact solution of TMG results in a
relationship between these scale factors $\lambda_i$, the cosmological
constant $ \lambda $   and the TMG coupling constant  $  \mu  $.

   We define triad basis 1-forms
\begin{equation}
l=\frac{1}{\sqrt{2}} \left( \omega^{0} - \omega^{1} \right) ,   \hspace{1cm}
n=\frac{1}{\sqrt{2}} \left( \omega^{0} + \omega^{1} \right) ,   \hspace{1cm}
m=\omega^{2}
\end{equation}
and taking their exterior derivative we have
\begin{equation}    \begin{array}{lll}
 dl & = & \frac{\textstyle{1}}{\textstyle{
 2\,\lambda_{0}  \lambda_{1} \lambda_{2} } } \left[
\left( \lambda_{0}^{2} + \lambda_{1}^{2}\right)  \,l \wedge m-
\left( \lambda_{0}^{2} - \lambda_{1}^{2}\right)  \,n \wedge m\right]  \\[3mm]
   dn & = &  \frac{\textstyle{1}}{\textstyle{
   2\,\lambda_{0}  \lambda_{1} \lambda_{2} } } \left[
\left( \lambda_{0}^{2} - \lambda_{1}^{2}\right)  \,l \wedge m-
\left( \lambda_{0}^{2} + \lambda_{1}^{2}\right)  \,n \wedge m \right] \\[3mm]
   dm & = & \frac{\textstyle{\lambda_{2}}}
   {\textstyle{\lambda_{0} \lambda_{1}} }\,l\wedge n
\end{array}
\end{equation}
Comparison of these expressions with eqs.(\ref{dlnm}) yields the Ricci
rotation coefficients
\begin{eqnarray}
\epsilon & = & \epsilon' = \sigma = \sigma' =0 , \nonumber
\label{hrotcoef}       \\
\tau & = & -\tau' =  -\frac{\lambda_{2}}{\lambda_{0} \lambda_{1}} ,
\nonumber \\
\kappa & = & -\kappa' = \frac{1}{2\,\lambda_{0}  \lambda_{1} \lambda_{2}}
\left(\lambda_{0}^{2} - \lambda_{1}^{2}\right)    ,          \\
\alpha & = & \frac{1}{2\,\lambda_{0}  \lambda_{1} \lambda_{2}}
\left( \lambda_{0}^{2} + \lambda_{1}^{2} -\lambda_{2}^{2} \right) . \nonumber
\end{eqnarray}
Inserting these quantities into eq.(\ref{lambda}) we obtain
\begin{equation}
  2\alpha \tau  +\kappa^{2}  -\tau^{2} = 3 \lambda  ,
\end{equation}
or using eq.(\ref{hrotcoef}) we have the explicit form
\begin{equation}
 ( \lambda_{0}+
\lambda_{1} +  \lambda_{2} )  ( - \lambda_{0}+ \lambda_{1} +  \lambda_{2} )
( \lambda_{0} - \lambda_{1} +  \lambda_{2} )  ( - \lambda_{0} - \lambda_{1}
+  \lambda_{2} ) = 12 \lambda
( \lambda_{0} \lambda_{1} \lambda_{2} )^{2} .
\label{rlambda}
\end{equation}
With the rotation coefficients (\ref{hrotcoef}) the TMG field equations
are drastically simplified and we have only two independent equations
\begin{equation}
3 \kappa \Phi_{11} + ( 2 \alpha - \tau ) \Phi_{00} = \mu \Phi_{00}
\label{heq1}
\end{equation}
\begin{equation}
3 \tau' \Phi_{11} + \kappa \Phi_{22} =
\frac{1}{2} \mu ( \lambda + \Phi_{02} )
\label{heq2}
\end{equation}
Comparison of these two equations and taking into account
eq.(\ref{lambda}) we obtain
a polynomial constraint involving only the scalar factors $
\lambda_{0},\, \lambda_{1},\, \lambda_{2} $.
This can be readily factorized to yield
\begin{eqnarray}
\frac{1}{( \lambda_{0} \lambda_{1} \lambda_{2} )^{5}}
( \lambda_{1}^{2} - \lambda_{0}^{2} )  ( \lambda_{0}^{2} - \lambda_{2}^{2} )
( \lambda_{1}^{2} - \lambda_{2}^{2} )
 ( \lambda_{0}+
\lambda_{1} +  \lambda_{2} )  ( - \lambda_{0}+ \lambda_{1} +  \lambda_{2} )
\nonumber \\
( \lambda_{0} - \lambda_{1} +  \lambda_{2} )  ( - \lambda_{0} - \lambda_{1}
+  \lambda_{2} ) = 0 .
\label{poly}
\end{eqnarray}
and the solution of the TMG field equations is now immediate.
When two of these scale factors coincide, say
$  \lambda_{1} =  \lambda_{2}  $, the simultaneous solution
of eqs.(\ref{rlambda}), (\ref{heq1}) and (\ref{heq2}) has either the form
\begin{equation}
\lambda_{0} = \lambda_{1} = \lambda_{2}= \frac{1}{2 \sqrt{ - \lambda}},
\label{bh1}
\end{equation}
or
\begin{equation}
\lambda_{0}= \frac{6 \mu}{ \mu^{2} -27 \lambda}    \hspace{2cm}
\lambda_{1} = \lambda_{2} =  \frac{3}{\sqrt{\mu^{2} -27 \lambda}}  .
\label{bh2}
\end{equation}
Clearly the solution in eqs.(\ref{bh1}) gives rise to the vanishing of
the Cotton tensor which is a trival solution of TMG.
The solution (\ref{bh2}) is a generalization of the
Vuorio solution \cite{v}  with a cosmological constant \cite{n}.
The general Bianchi Type $VIII$ solution where
the scale factors are related by $\lambda_0 \pm
\lambda_1 \pm \lambda_2 = 0$ does not admit a cosmological constant.
In all cases for the DJT coupling constant we get\footnote{We take
this opportunity to correct an extra factor  of $  \; 2  $  that crept
into the denominator of this expression in \cite{nb}.}
\begin{equation}
\mu =  \frac{\lambda_{0}^{2} + \lambda_{1}^{2} + \lambda_{2}^{2}}
{\lambda_{0}  \lambda_{1} \lambda_{2}}
\label{scale}
\end{equation}
from eqs.(\ref{heq1}), (\ref{heq2}).

\section{Black holes in TMG}

     The Bianchi Type $VIII$ homogeneous space-time can be re-identified to
yield a black hole solution of TMG in the same way that AdS has been
re-identified by Ba\~{n}ados, Teitelboim and Zanelli
to yield the black hole solution of the 3-dimensional Einstein
field equations with a cosmological constant \cite{btz}.
This process was further clarified by Horowitz and Welch \cite{hw}
and we shall use their approach here. Earlier  \cite{n}, such a
re-identification was carried out for $\lambda_1 = \lambda_2 $
which results in a stationary 2-papameter black hole solution.
But now we shall show that
the general Bianchi Type $VIII$ solution can also be re-identified to
yield a non-stationary, non-axisymmetric 3-parameter solution that
has many properties akin to those of a black hole.

     For the general case of Bianchi Type $VIII$ solution we shall
use introduce a parameter $p$
\begin{equation}
 \lambda_{1} = \frac{3+p^2}{\mu (1-p)}
\;\;\;\;\;\;
 \lambda_{2} = \frac{3+p^2}{\mu (1+p)}
\end{equation}
and use the root
\begin{equation}
\lambda_{0} =  \lambda_{1} +  \lambda_{2} = \frac{2 (3+p^2)}{\mu (1-p^2)}
\end{equation}
of eq.(\ref{rlambda}). Then the metric is given by
\begin{equation}   \begin{array}{ll}
d s^2 =
& \mu^2          \left(
\frac{\textstyle{1-p^2}}{\textstyle{3+p^2}} \right)^2
                       \left\{
4 \left( d \psi + \cosh \theta d \phi \right)^2
- \left( 1 + p^2 - 2 p \, \cos 2 \psi \right) d \theta^2
 \right.    \\[5mm]
 & + 4 p \, \sin 2 \psi \, \sinh \theta \; d \theta \, d \phi
- \left( 1 + p^2 + 2 p \, \cos 2 \psi \right) \sinh^2 \theta \, d \phi^2
           \left.            \right\}
\end{array}
\label{met1}
\end{equation}
and introducing a new variable
\begin{equation}               \begin{array}{rl}
\sinh \left(\frac{\textstyle{\theta}}{\textstyle{2}} \right) = &
\frac{\textstyle{\mu}}{\textstyle{2}} \,
\frac{\textstyle{1-p^2}}{\textstyle{3 + p^2}} \, R \\[3mm]
\end{array}
\label{hw1}
\end{equation}
we find that the Bianchi $VIII$ metric (\ref{met1}) becomes
\begin{equation}   \begin{array}{ll}
 d s^2 =
& \frac{\textstyle{4}}{\textstyle{\mu^2}}
\left( \frac{\textstyle{3+p^2}}{\textstyle{1-p^2}} \right)^2
\left\{ d \psi
+ \left[ 1+ \frac{\textstyle{1}}{\textstyle{2}}
 \left(\frac{\textstyle{ 1-p^2}}{\textstyle{ 3 + p^2}} \right)^2
\mu^2 R^2 \right] d\phi \right\}^2  \\[9mm]
& -\frac{\textstyle{( 1 + p^2 - 2 p \, \cos 2 \psi ) }}
 {\textstyle{ 1
 + \frac{\textstyle{1}}{\textstyle{4}}
\left( \frac{\textstyle{1-p^2}}{\textstyle{3+p^2}} \right)^2 \mu^2 R^2 }}
 d R^2 + 4 p \, \sin 2 \psi \, R \; d R \, d \phi \\[9mm]
& -  \left( 1 + p^2 + 2 p \, \cos 2 \psi \right)
 \left[ 1 + \frac{\textstyle{1}}{\textstyle{4}}
\left( \frac{\textstyle{1-p^2}}{\textstyle{3+p^2}} \right)^2
\mu^2 R^2  \right]  R^2  d \phi^2 \\[5mm]
\end{array}
\label{met2}
\end{equation}
which is a limiting form of the black hole metric. We can dress up
this metric with two parameters through the transformations \cite{hw}
\begin{eqnarray}
\left(  \frac{\textstyle{\mu}}{\textstyle{2}}    \,
 \frac{\textstyle{1-p^2}}{\textstyle{3+p^2}}
 \right)^2  R^2
& = & \frac{r^2 - r_{+}^{\;2} }
 {r_{+}^{\;2} - r_{-}^{\;2} }    \nonumber \\ [5mm]
 \psi & = & F(t,\varphi)   \\ [5mm]
 \phi & = & \gamma \,\varphi            \nonumber 
\label{hw2}
\end{eqnarray}
where $ r_{+} $ , \,\,  $ r_{-} $  and  $ \gamma $ are constants.
The proper choice of $F$ that will enable us to interpret $\varphi$
as an angle is determined from the solution of
\begin{equation}
\left( \frac{\partial F}{\partial \varphi} - \frac{M}{6} \,
\sqrt{ \frac{6}{6aJ -M}} \right)^2 \,
= \,  \frac{3}{8} \, \frac{ a^2 \,J^2}{6aJ-M} \,  ( 1 + p^2 + 2 p \cos 2 F )
\label{eqf}
\end{equation}
where
\begin{equation}
a= \frac{1-p^2}{3+p^2} \, \, , \hspace{5mm}
 M=\frac{3}{2}a^2 \mu^2 \left( r_{+}^{2} + r_{-}^{2} \right),  \hspace{5mm}
 J=  a  \mu^2  \, r_{+} r_{-}
\end{equation}
and we shall find it convenient to define
\begin{equation}
f(t,\phi) = \frac{2}{a \mu} \, \sqrt{ \frac{6}{6aJ -M}} \,\,
\frac{\partial F}{\partial t} .
\end{equation}
Note that $F$ is determined from eq.(\ref{eqf}) up to an arbitrary
additive function of $t$ alone which can be fixed by physical requirements.
Then the metric (\ref{met2}) takes the form
\begin{equation}   \begin{array}{ll}
d s^2 =
& \frac{1}{6} \left( 6aJ-M \right) f^2 d t^2 + 2 f\left( a \mu r^2
- \frac{\textstyle{J}}{\textstyle{2\mu}} \sqrt{B} \right)
          d t \, d\varphi \\[7mm]
& - \frac{\textstyle{ 1 + p^2  - 2 p \, \cos 2 F}}
{\textstyle{\frac{\textstyle{1}}{\textstyle{4}} a^2 \mu^2 r^2
- \frac{\textstyle{1}}{\textstyle{6}} M +
\frac{\textstyle{J^2}}{\textstyle{4\mu^2 r^2} } }}\, d r^2
+  4 p \, \sqrt{\frac{\textstyle{6}}{\textstyle{6aJ -M}}}
\, \sin 2 F \;\, r \; d r \, d \varphi               \\[11mm]
&- \frac{\textstyle{1}}{\textstyle{6aJ-M}}
\left[ \left(6aJ-M\sqrt{B} \right) \sqrt{B} -\frac{3}{2}
a^2 \mu^2 r^2 \left( 4 - B \right) \right] r^2 d \varphi^2
\end{array}
\label{met3}
\end{equation}
with the definition
\begin{equation}
B = 1+ p^2 + 2p \cos2 F .
\end{equation}
It is seen that the constants $  r_{+} $ and  $  r_{-} $ introduced above
are given by
\begin{equation}
 r_{\pm}^{2} = \frac{M}{3 a^2 \mu^2}\left[ 1 \pm \left( 1
 - 9 \frac{a^2 \, J^2}{M^2} \right)^ \frac{1}{2}\right]
\end{equation}
describe the location of the outer and inner event horizons

     If we further specialize to $p=0$ which is equivalent to
$\lambda_1 = \lambda_2 $, the metric becomes stationary \cite{n}.

\section{Tele-parallelism}

     The solution of TMG that we have discussed above has a very interesting
property that was not noted before. It describes a parallelizable manifold
where we can introduce a global frame so that the curvature can be made
to vanish at the expense of introducing torsion.
The notion of a parallelizable manifold is due to Cartan and Schouten
\cite{cartsch} and the simplest example they chose
as an illustration of their ideas was $S^3$ as a parallelizable manifold.
We shall now show that the vacuum Bianchi Type $VIII - IX$ exact solutions
of TMG provide new examples of a parallelizable 3-dimensional manifolds.
They are therefore of interest from a purely mathematical point of
view as well.

     In order to facilitate comparison with Cartan and Schouten's
treatment of $S^3$ as much as possible, we shall now use
a triad with 2 space-like legs and a third
leg which can be either time-like or space-like depending on a
parameter $\epsilon = \mp 1$. Thus we write the metric in the form
\begin{equation}
d s ^2 = \epsilon \, (\omega^0)^{\;2}  + (\omega^1)^{\;2}
        + (\omega^2)^{\;2}
\end{equation}
where the co-frame is given by eqs.(\ref{homframe}) and depending
on $\epsilon$ the $\sigma^i$ are now the left-invariant 1-forms of
either Bianchi Type $VIII$, or $IX$. We shall hence-forth write
$S^3$ in quotation marks as our treatment will also cover the
case of Lorentz signature. Left-invariant 1-forms $\sigma^i$ satisfy
the Maurer-Cartan equations (\ref{maurerc}) with structure constants
\begin{equation}
 C_{\;\;12}^{0}= \epsilon, \hspace{1cm}   C_{\;\;20}^{1}= 1, \hspace{1cm}
 C_{\;\;01}^{2}= 1                        .
\label{struconst}
\end{equation}
that have a dependence on $\epsilon$ to match the signature of space-time.
Note that for Bianchi Type $IX$ we shall require the satisfaction of the
Euclidean TMG field equations. Then starting with the frame (\ref{homframe})
we find the connection 1-forms
\begin{equation}         \begin{array}{lll}
\omega^{1}_{\;\;0} & = & \frac{\textstyle{\epsilon}}
  {\textstyle{ 2 \lambda_0  \lambda_1  \lambda_2 } }
( \lambda_{0}^{\;2} + \lambda_{1}^{\;2} - \lambda_{2}^{\;2} ) \, \omega^2
                            \\[3mm]
\omega^{2}_{\;\;0} & = & \frac{\textstyle{1}}
  {\textstyle{ 2 \lambda_0  \lambda_1  \lambda_2 } }
( \lambda_{0}^{\;2} - \lambda_{1}^{\;2} + \lambda_{2}^{\;2} ) \, \omega^1
                            \\[3mm]
\omega^{1}_{\;\;2} & = & \frac{\textstyle{1}}
  {\textstyle{ 2 \lambda_0  \lambda_1  \lambda_2 } }
( \lambda_{0}^{\;2} + \lambda_{1}^{\;2} - \lambda_{2}^{\;2} ) \, \omega^0
\end{array}
\label{constar}
\end{equation}
from Cartan's equations of structure.

     The crucial relation which makes the Bianchi Type $VIII - IX$ solutions
of TMG a parellelizable manifold is the relation
\begin{equation}
\epsilon_0 \lambda_0 + \epsilon_1 \lambda_1 + \epsilon_2 \lambda_2 = 0
\hspace{15mm} \epsilon_{i}^{\;2} = 1, \;\; i=0,1,2
\label{star}
\end{equation}
that is required for the vanishing of the curvature scalar.
This is also the condition for the embeddability
of the 3-manifold into 4-dimensional (anti)-de Sitter
universe, or the (pseudo)-sphere  \cite{oz}.
Now using eq.(\ref{star}) the connection 1-forms (\ref{constar}) reduce to
\begin{equation}         \begin{array}{lll}
\omega^{0}_{\;\;1} & = & - \epsilon  \, \epsilon_0 \epsilon_1 \,  \sigma^2 ,
                        \\[2mm]
\omega^{2}_{\;\;0} & = &  - \epsilon_0 \epsilon_1 \, \sigma^1 , \\[2mm]
\omega^{1}_{\;\;2} & = & - \epsilon_0 \epsilon_1  \, \sigma^0   ,
\end{array}
\label{constar2}
\end{equation}
which up to a harmless looking, but nevertheless very important factor
of $2$ are the same as the corresponding results for $`` S^3 "$.
Similarly the curvature 2-forms are simplified drastically
\begin{equation}         \begin{array}{lll}
\theta^{1}_{\;\;2} & = & 2 \, \epsilon  \epsilon_1 \epsilon_2 \,
 \sigma^2 \wedge \sigma^1 , \\[2mm]
\theta^{2}_{\;\;0} & = & 2 \epsilon_0 \epsilon_2 \,\sigma^0 \wedge \sigma^2
,  \\[2mm]
\theta^{1}_{\;\;0} & = & 2 \epsilon_0 \epsilon_1 \, \sigma^0 \wedge \sigma^1
\end{array}
\label{curvstar2}
\end{equation}
which are identical to their respective expressions for $`` S^3 "$
up to a factor of $8$.

   Following Cartan and Schouten we now introduce the contorsion 1-forms
$ K^{i}_{\;k} $ so that the full connection is given by
\begin{equation}
\tilde{\omega}^{i}_{\;k} = \omega^{i}_{\;k} + K^{i}_{\;k}
\label{contort}
\end{equation}
where we make the usual {\it Ansatz}
\begin{equation}
 K^{i}_{\;k}  = \nu \, \omega^{i}_{\;k}
\label{concon}
\end{equation}
with $\nu$ a constant. The condition for the vanishing of the full
curvature
\begin{equation}
\tilde{\theta}^{i}_{\;k} = d \tilde{\omega}^{i}_{\;k} +
\tilde{\omega}^{i}_{\;m} \wedge \tilde{\omega}^{m}_{\;k} = 0
\label{curv0}
\end{equation}
has the trivial solution $\nu = -1$ but there is also
the non-trivial solution
\begin{equation}
\nu = -2
\label{lamm2}
\end{equation}
which we shall hence-forth adopt. For $`` S^3 "$ we had $+1$ for the
non-trivial value of $\nu$. The torsion 2-form is given by
\begin{equation}
T^{i}  =  K^{i}_{\;k}  \wedge \omega^{k}
\label{torsion}
\end{equation}
and with $\nu = - 2$ we get
\begin{equation}         \begin{array}{lll}
T^{0} & = & - 2 \epsilon  \, \lambda_0  \, \sigma^1 \wedge \sigma^2 , \\[1mm]
T^{1} & = & - 2 \lambda_1  \, \sigma^2 \wedge \sigma^0 , \\[1mm]
T^{2} & = & - 2 \lambda_2  \, \sigma^0 \wedge \sigma^1
\end{array}
\label{torsion2}
\end{equation}
which differ from the torsion 2-forms for $`` S^3 "$ by a factor of $2$
after allowing for the identity of the scale factors.
It is straight forward to verify that the covariant exterior
derivative of the torsion 2-form with respect to the full connection
$\tilde{D}$ vanishes
\begin{equation}
\tilde{D} T^{i} =  0
\label{covtorsion}
\end{equation}
which, in view of eq.(\ref{curv0}), is the first Bianchi identity.

     It remains to verify that the Cartan-Schouten equations are satisfied
for the Bianchi Type $VIII$ and $IX$ solutions.
We recall that for $`` S^3 "$ these equations are given by
\begin{equation}
 T_{i} \wedge ^*\hspace{-1mm}T_{k} = g_{ik} \; ^*1
\label{cs1}
\end{equation}
where $^*1$ denotes the volume 3-form and
\begin{equation}
 \omega^i \wedge T_i = - 4 \epsilon l^2
K^{i}_{\;\,j} \wedge K^{j}_{\;\,k} \wedge K^{k}_{\;\,i}
\label{cs2}
\end{equation}
respectively.

   For the first Cartan-Schouten equation the check that
$ T_{i} \wedge ^*T_{k} = 0 $  for $i \ne k $ is immediate
from eqs.(\ref{torsion2}). For the verification of the diagonal components
we have from eqs.(\ref{torsion2})
$$ \begin{array}{ccc}
T_{0} =  2 \lambda_0 \sigma^1 \wedge \sigma^2 &
\hspace{1cm} \Longrightarrow \hspace{1cm}      &
^*T_{0} = 2 \epsilon \lambda_1 \lambda_2   \sigma^0  , \\
T_{1} =  2 \lambda_1 \sigma^2 \wedge \sigma^0  &
\hspace{1cm} \Longrightarrow \hspace{1cm}        &
^*T_{1} = 2 \lambda_0 \lambda_2   \sigma^1 \end{array}  $$
see \cite{fl} and similarly for the last component. Thus we have
\begin{equation}
\frac{1}{4} T_{i} \wedge ^*\hspace{-1mm}T_{k} =  g_{ik} \; ^*1
\label{cs19}
\end{equation}
in place of eq.(\ref{cs1}).
Finally a straight-forward calculation shows that
\begin{equation}
 \omega^i \wedge T_i =  2 \mu \; ^{*}1 =
 \frac{\epsilon}{24} \mu \lambda_0 \lambda_1 \lambda_2
      K^{i}_{\;\,j} \wedge K^{j}_{\;\,k} \wedge K^{k}_{\;\,i}
\label{cs22}
\end{equation}
which up to a constant of proportionality agrees with eq.(\ref{cs2}).

     The starting point in the work of Chern and Simons \cite{chs} is
that all oriented 3-manifolds are parallelizable. The gravitational
Chern-Simons term is the Cotton tensor which vanishes for  $S^3$,
the classic example of a parallelizable manifold.
We have shown that the Bianchi Type $VIII - IX$ exact solutions of
TMG are also orientable and parallelizable as they satisfy the
conditions (\ref{curv0}) for zero-curvature and the Cartan-Schouten
equations up to simple modifications of the factors of proportionality.
These are new examples of 3-dimensional parallelizable manifolds.

\section{Acknowledgements}

   One of us (Y.N.) thanks T. Dereli for an interesting discussion
on ref. \cite{cartsch}. This work was in part supported by Turkish
Academy of Sciences.

\end{document}